\documentclass[pra, amsmath,amssymb,aps,10pt,twocolumn,superscriptaddress]{revtex4-1}
\usepackage{graphicx}
\usepackage[usenames, dvipsnames]{color}
\usepackage{dcolumn}
\usepackage{bm}
\usepackage{color}
\usepackage[colorlinks,bookmarks=false,citecolor=blue,linkcolor=blue,urlcolor=black]{hyperref}
\usepackage{ulem}

\begin{document}

\preprint{APS/123-QED}


\title{Double ionization of a three-electron atom: Spin correlation effects}
\author{Dmitry K. Efimov} \email{dmitry.efimov@uj.edu.pl}
 
\affiliation{Instytut Fizyki imienia Mariana Smoluchowskiego, Uniwersytet Jagiello\'nski, \L{}ojasiewicza 11, 30-348 Krak\'ow, Poland}
\author{Jakub S. Prauzner-Bechcicki}
 \affiliation{Instytut Fizyki imienia Mariana Smoluchowskiego, Uniwersytet Jagiello\'nski, \L{}ojasiewicza 11, 30-348 Krak\'ow, Poland}
\author{ Jan H. Thiede}
  \affiliation{Philipps-University Marburg, Biegenstra{\ss}e 10, 35037 Marburg, Germany}
\author{Bruno Eckhardt}  \thanks{$\dagger${Deceased 7 August 2019.}}
 \affiliation{Philipps-University Marburg, Biegenstra{\ss}e 10, 35037 Marburg, Germany}
 \author{Jakub Zakrzewski}
 \affiliation{Instytut Fizyki imienia Mariana Smoluchowskiego, Uniwersytet Jagiello\'nski, \L{}ojasiewicza 11, 30-348 Krak\'ow, Poland}
\affiliation{Mark Kac Complex Systems Research Center, Jagiellonian University, \L{}ojasiewicza 11, 30-348 Krak\'ow, Poland}
\date{\today}

\begin{abstract}
We study the effects of spin degrees of freedom and wave function symmetries 
on double ionization in three-electron systems.
Each electron is assigned one spatial degree of freedom. The resulting three-dimensional Schr\"odinger equation 
is integrated numerically using grid-based Fourier transforms. We reveal three-electron effects on the double ionization yield by comparing signals for different ionization channels. We explain our findings by the existence of fundamental differences between three-electronic and truly two-electronic spin-resolved ionization schemes.
We find, for instance, that double ionization from a three-electron system is dominated by electrons that have the opposite spin. 
\end{abstract}

\maketitle


	\section{\label{sec:introduction}Introduction}

Approaching the attosecond time scale in experiments allows one to directly probe internal atomic dynamics.
The improvement in experimental techniques encourages one to develop new and refine existing  theoretical tools in a way that will advance our understanding of processes and correlations inside atoms and molecules that take place under strong electromagnetic field irradiation. That task remains 
at the focus of current strong-field physics.

Correlations between electrons in atoms and molecules 
and their interaction with femtosecond pulses have been studied extensively \cite{Fittinghoff92,Schulz2004,Walker94}. The range 
and methods of these studies depended on how an electronic correlation was understood by researchers. For example, a theoretical definition of an electronic correlation was introduced in \cite{Grobe94,Byczuk12}, but its correspondence with experimentally measurable quantities stays unclear. On the other hand, there are many experimentally accessible signatures of correlations. In the context of multiple ionization these are 
a characteristic ``knee'' in the field amplitude dependent 
double ionization yield \cite{Kondo93,Walker94} and the shape of two-electron momentum distributions \cite{Goreslavskii01,Pfeiffer11a,Lein00,Moshammer02,Bergues12,Chen17}. The former was explained by 
the presence
of two double ionization channels: sequential double ionization (SDI), when electrons are released without experiencing a dynamical interaction, and non-sequential double ionization  (NSDI), for which electronic interactions play a decisive role, for example by means of an electron--parent-ion recollision process \cite{Faisal97,Becker99,Mauger10,Watson97}. In the electron momenta distribution, the pronounced ``fingerlike'' structure originates in Coulomb correlations \cite{Parker06,rudenko2007correlated,staudte2007binary,Zielinski16,Emmanouilidou08,Ye08}. Additionally, angular distributions of electronic momenta possess a correlation fingerprints as well \cite{Jiang13}. All these observations are nicely reviewed in 
\cite{Faria11,becker2012theories}. Momentum distributions also allow one to test how electronic correlations affect the delay between two electron ionization times \cite{Pfeiffer11b,Emmanouilidou15}. Finally, multielectronic correlations can manifest themselves in High Harmonic Generation (HHG) \cite{Sukiasyan09,Smirnova09,Prager01,Telnov09} and affect sub-barrier strong-field tunneling times \cite{Ivanov14multielectron}. For instance, Coloumb correlations prevent simultaneous rescattering of ionizing electrons from the ion they left behind,
thus shifting the second plateau cutoff position of high harmonic spectra \cite{koval2007nonsequential}. 

From a theoretical point of view, correlations of electrons may be revealed by the examination of subsystem dynamics. For example, in \cite{Ivanov14multielectron,Ivanov18} multielectron correlations are considered to affect the single-electron ionization and HHG. The electron can also experience simultaneous rescattering on mixed neutral and ionized states and thus influence generation of harmonics \cite{Li19,Tikhomirov17}. Chattopadhyay and Madsen \cite{chattopadhyay2018electron} consider a two-electron correlation effect on a single ionization of diatomic  molecules. 

A minimal system where possible effects due to third active electron on double ionization may be examined has just three electrons, such as Li. An early seminal study of Li atom ionization demonstrated also the necessity to carefully take into account the electron spin \cite{Ruiz05}. 
This is to be contrasted with standard two-active electrons treatments that implicitly assume a spatially symmetric wave function (corresponding to antisymmetric configuration of spins). The effects resulting from three-electron spin configuration we shall call spin correlation effects. While \cite{Ruiz05} considered the ionization of lithium with few large frequency photons, the progress in numerical methods allows us nowadays to treat three active electrons at optical frequencies \cite{Thiede18}, though within a reduced dimensionality model.

The spin correlation effects are to be distinguished from the Coulomb electron-electron correlation and in particular from Coulomb interaction of the third electron with the remaining pair. The latter is not an easy object to study as any arbitrary changes in the internal atomic potential yield the considerable shifts of ionization potential values and thus of Keldysh parameter values \cite{Eberly00}.

In the present paper we employ our three-active electrons model \cite{Thiede18} to study effects of three-electron correlations on the double ionization process.
In the pure 3-electron atom, Li, the three electrons are not symmetric, since there is 
a single weakly bound p electron 
and two strongly bound 1s$^2$ electrons.
While Li ionization potentials are unfavorable for studying multiphoton correlation effects \cite{Ruiz05}, to identify the effects of correlations, it 
seems better
to consider  
atoms
like boron or aluminum (with ns$^2$np$^1$ electrons forming the outer shell). 
Even this is highly difficult as a real multiphoton regime
would require going to very low frequencies. To stay with standard optical frequencies, say $\omega=0.06$ a.u.
which corresponds 
to about 760 nm wavelength of laser irradiation, 
we consider instead an artificial 3-electron atom
with single, double and triple ionization thresholds corresponding to Ne. 
To be closer to experiments, 
we simulate ionization dynamics under the influence of experimentally achievable 5-cycles-long laser pulses,
rather than the very short 2-cycle pulse we considered previously \cite{Thiede18}.

The correlation effects are revealed via comparing results of strong-field ionization simulation for 3-electron and 2-electron models of the same atom. The latter is specially defined to be in a good correspondence with the former, preserving space geometry and double ionization potential values. The spin correlation effects study implies keeping interaction terms in two- and three- potentials as close to each other as possible.

The paper is organized as follows. First we describe in Section \ref{sec:models} models of two- and tree-electron atoms we use, underlining physical differences between them that can not be eliminated by simple changes of parameters. 
We then discuss in Section \ref{sec:pathways} the different ionization processes in the presence
of spin degree of freedom. In Section \ref{sec:results_3e} 
we show how such differences manifest themselves in the output of atomic ionization simulations. 
We
 close with a summary and conclusions in Section \ref{sec:conclusion} . 
Atomic units are used throughout this paper unless stated otherwise. For the sake of clarity, we note that 1 a.u. of energy is equal to 27.2 eV; at the same time 0.1 a.u. of electric field corresponds to $3.5\cdot 10^{14}$ W/cm$^2$ of laser intensity.

	\section{\label{sec:models}{Simplified atomic models}}
A numerical calculation of the dynamics of three electrons in the full space is still beyond the reach of current numerical capabilities. We introduce, therefore, models where each electron is restricted to move along one-dimensional track so that the dimensionality of position space does not exceed three. We first construct the three-active electrons model and then define the two active electrons restricted model.

\begin{figure}
	\includegraphics[width=1.0\linewidth]{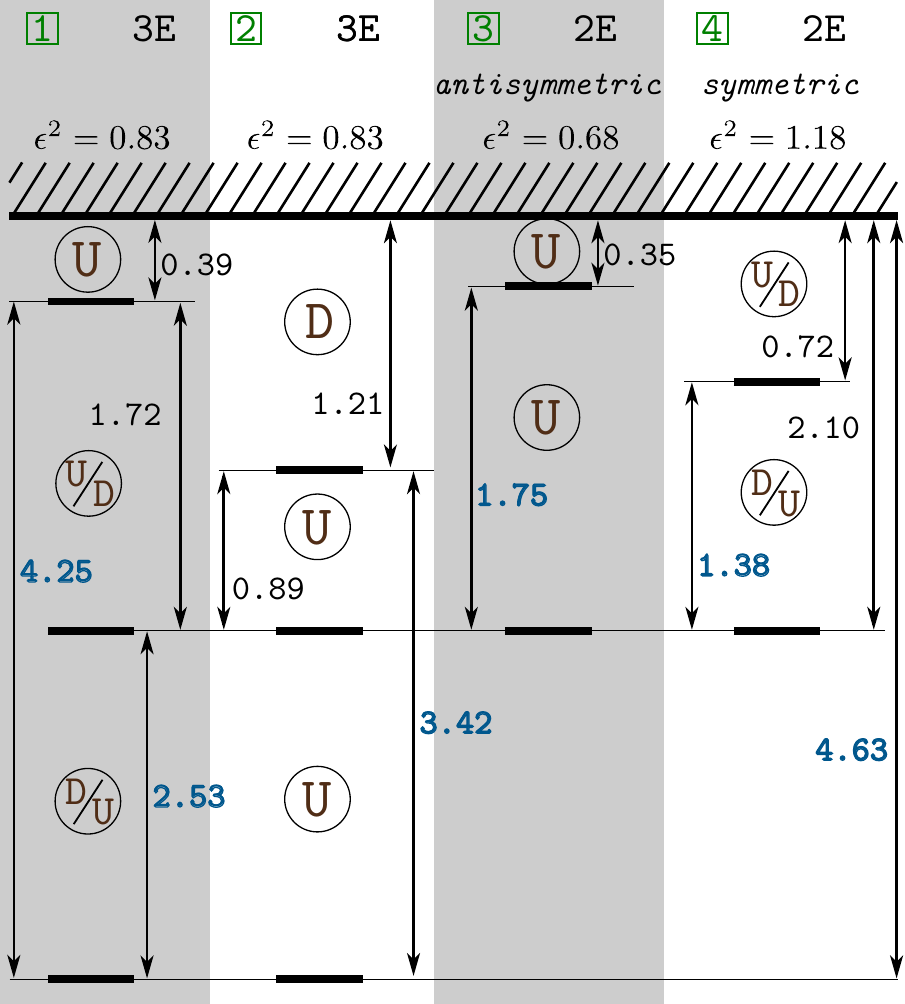}
	\caption{ (Color online) Level schemes for different ionization scenarios (marked with numbers in green squares). Ionization events are visualized  from top to the bottom. Scenarios 1 and 2 correspond to an ionization in the three-active electrons model with potential (\ref{ham3e}), softening parameter $\epsilon=\sqrt{0.83}$ a.u.: in scenario 1 the electron ionized first is a (U) electron, whereas in scenario 2 it is the (D) electron. The notation (U/D) vs (D/U) for the second and third ionization events refers to both possibilities (U) before (D) and (D) before (U), respectively. Scenarios 3 and 4 both correspond to an ionization in the two-active electrons model with the potential (\ref{ham2e}). In scenario 3 the antisymmetric configuration (both spins up) with $\epsilon_{antisym}=\sqrt{0.68}$ a.u. is considered, whereas scenario 4 corresponds to a symmetric configuration (U) vs (D) or (D) vs (U) with $\epsilon_{sym}=\sqrt{1.18}$ a.u. Letters in circles denote the spin of the electron ionized in a particular event. Numbers reflect the ionization potential values: the bold blue ones are results of numerical simulations while the black ones are obtained as simple differences. }
	\label{fig1}
\end{figure}

 \noindent\paragraph{\label{sec:3emod}Three-active electrons model.} In this model, 
 {the motion of each electron is restricted to a 
 one-dimensional line, forming }
 (i) an angle of $\pi/6$ between each other and (ii) an angle $\gamma$ ($\tan \gamma = \sqrt{2/3}$) with the electric field polarization direction. We should point out here that the vector of electric field is not lying in any of the two-electronic planes, but forms equal angles with them. This unusual configuration is identified on the basis of an adiabatic analysis that assumes that the ionization process is most effective along the lines defined by the saddles of the potential energy in the presence of the instantaneous static electric field \cite{sacha2001triple}. These saddles can be considered as transition states leading to efficient channels for ionization. As the field amplitude changes during the pulse, these saddles move along lines inclined at constant angles with respect to the field polarization axis and to each other in a fixed 
{configuration, independent of field strength} \cite{sacha2001triple}. Those lines are then taken as tracks to which each electron's motion is confined. The Hamiltonian of three-electron system in the discussed geometry then reads
\begin{equation}
H=\sum_{i=1}^3\frac{p_i^2}{2}+V(r_1,r_2,r_3)
\label{ham3e}
\end{equation}
with 
\begin{eqnarray}
V(r_1,r_2,r_3)&=&-\sum_{i=1}^3\left(\frac{3}{\sqrt{r_i^2+\epsilon^2}} +\sqrt{\frac{2}{3}}F(t)r_i \right) \nonumber \\
&+&\sum_{i,j=1 i<j}^3\frac{1}{\sqrt{(r_i-r_j)^2+r_ir_j+\epsilon^2}},
\label{pot3e_std}
\end{eqnarray}
%
where $r_i$ and $p_i$ correspond to the $i$-th electron's coordinate and momentum, respectively, $\epsilon$ is a parameter softening the Coulomb singularity 
and $F(t) = -\partial A/\partial t$ denotes time dependent field with the vector potential
\begin{equation}
A(t) = \frac{F_0}{\omega_0} \sin^2 \left( \frac{\pi t}{T_p} \right) \sin(\omega_0 t), \quad 0<t<T_p.
\label{pulse}
\end{equation}
Here, the pulse length $T_p = 2\pi n_c /\omega_0$ is 
{taken to be a multiple} of the number of cycles $n_c$. The corresponding ionization energies are presented in Fig. \ref{fig1} -- the softening parameter is chosen to give a ground state energy equal to the triple ionization potential of Ne, i.e.  $\epsilon=\sqrt{0.83}$ a.u. so that 
$I_p=4.63$ a.u. i.e. 126 eV. Time-dependent Schr\"oodinger equation is solved on a spatial grid with the use of split operator technique and Fast Fourier transform with algorithms described in details elsewhere~\cite{Thiede18,prauzner2008quantum}. The largest grid used, having 2048 points in each direction covering 400 a.u. of physical coordinate space, required about a week of 192 cores for 5-cycles pulse evaluation. The initial state was found by imaginary time propagation in an appropriate symmetry subspace for much smaller grid involving, typically, 512 points in each
direction corresponding to 100 a.u. The ionization yields are calculated by integrating the electronic fluxes through different ionization regions borders, which in turn are expressed as a surface integrals of probability currents calculated directly from the wavefunction and its gradient. The procedure is described in detail in \cite{Thiede18}. We use absorbing boundary conditions at edges of the integration box~\cite{prauzner2008quantum}.

\noindent\paragraph{Two-active electrons model.}
The two-active electrons model is built consistently from the three electrons model discussed above. We restrict the electronic motion to one-dimensional tracks that form a plane and cross at angle $\pi/6$ as in the three electrons model. The same two-electron model was used in~\cite{prauzner2008quantum,prauzner2007time,eckhardt2008,eckhardt2010phase,Efimov18}. In the 2-dimensional geometry the electric field vector is forced to lay in that plane -- in contrast to the three electrons case, thus forming a different angle $\cos (\pi/6) = \sqrt{3}/2$ with electronic axes. For the sake of comparison between both models, we impose the electric field operator geometrical prefactors to be the same and equal to $\sqrt{2/3}$, as introduced earlier in (\ref{ham3e}). The two-electron Hamiltonian then reads

\begin{multline}
	H = \sum_{i=1}^{2} \left( \frac{p_i^2}{2} - \frac{Z}{\sqrt{r_i^2 +\epsilon^2}} +  \sqrt{\frac{2}{3}}F(t)r_i \right) + \\ \frac{1}{\sqrt{(r_1-r_2)^2 + r_1 r_2 + \epsilon^2}},
	\label{ham2e}
\end{multline}   
where $Z$ is a nuclear charge, set either as $Z=2$ (Ne atom) or $Z=3$ (Ne$^+$ ion). 
Two dimensional calculations have been performed on 2048 points grid corresponding to 400 a.u. per direction with initial state found on a smaller grid of 512 points.

\paragraph{Spin configurations.} 

In the case of three electrons the whole system evolution can be described by a single three-electronic spatial function (see Appendix \ref{app:wavefunction} for the discussion). Without loss of generality we may  consider a case with two spin-up (U) and one spin-down (D) electrons.
The spatial wave function is antisymmetric in the UU plane and symmetric in all other planes.

During double ionization (U) and (D) or (U) and (U) electrons can be ionized. To mimic this picture with two-active electrons model, one needs to consider two problems with different symmetries: spatially symmetric, corresponding to (U) and (D) ionized electrons, and  spatially antisymmetric, corresponding to (U) and (U) ionized electrons.

\paragraph{Calculation of ionization potentials.} 

The definition of atomic models immediately implies the way for calculating a set of ionization potentials. To find the ground state wave function and the corresponding eigenenergy we employ the imaginary-time evolution starting from a function with the required symmetry. Such a procedure works for one-, two- or three-electron functions. 

The three-electron model allows one to resolve ionization channels in respect to number, sequence and spin of ionized electrons (see the details in \cite{Thiede18}), and thus introduce different scenarios for discussed processes. Scenarios analyzed in the present study are presented in Sec.~\ref{sec:results_3e}. For the sake of clarity and completeness of methodology section, we shall expose procedure of computation of ionization potential discussing a particular example of an ionization channel in the three-electron model.
Let the (U) electron be ionized first, and followed then by (D) and (U) electrons (the scenario 1 in Fig. \ref{fig1}). 
 The triple ionization potential 4.63 a.u. is defined by the ground state of the three-electron wavefunction. One can easily calculate the single ionization potential of a double ion ($Z=3$), i.e. when two electrons are already ionized. The double ion has just one electron, so the potential value 2.53 a.u. is spin-independent. The difference of it with the triple ionization potential gives the double ionization potential value of 2.10 a.u. The single ionization potential is calculated as the difference of triple ionization potential and double ionization potential 4.25 a.u. of a single ion ($Z=3$) possessing (U) and (D) electrons (and thus spatially symmetric) yielding 0.39 a.u.
 
 In the two-electron model getting ionization potentials is even simpler. First, one finds ground state of a two-electron system ($Z=2$), i.e. a double ionization potential. Then, the single ionization potential (ground state) of a single ion ($Z=2$) is calculated in corresponding one-electron system. Difference between these two ionization potentials gives the single ionization potential of the two-electron system.

	\section{\label{sec:pathways}Spins and ionization potentials}

In the system considered several ionization scenarios are possible, accounting for the different channels of single, double and triple ionization. In the following we will focus primarily on double ionization events, thus triple direct ionization is not discussed at all. Discussion of the triple ionization as described within our three-active electrons model can be found elsewhere~\cite{Thiede18}. Also, for the parameters used in our simulations, triple ionization yields are about 3 orders smaller in magnitude than those for double ionization, thus we can assume that the loss of electrons to triple ionization channel is negligible.

Generalizing the already mentioned spin independence of the Ne$^{++}$ ionization potential, one comes to an equality of double ionization (DI) potentials of $2.10$ a.u.
for all possible spin configurations of first two electrons, i.e. (DU), (UD) and (UU) (see Fig.~\ref{fig1}). Here (DU) denotes that the first of the escaping electrons has spin down (D) and the second has spin up (U), and 
similar for the other cases. These two electrons can escape either sequentially or simultaneously. Of course, for the simultaneous escape (DU) and (UD) configurations are indistinguishable. 
	
The value of the {single electron ionization (SI) potential}
 of an atom does depend, however, on the  ionization channel. If the first ionizing electron is (U) -- like in the example from the previous section -- then its release leads to the formation of a singly charged ion with a 
(UD) pair that is described by a spatially symmetric wave function. We shall call this path scenario 1 (see Fig.~\ref{fig1}). However, if the first ionizing electron is (D) then a (UU) pair is formed, with a spatially antisymmetric wave function. We shall call this path scenario 2 (see Fig.~\ref{fig1}).


One should note that the two (U) electrons are indistinguishable in the analyzed model. Therefore there are only two single ionization channels, i.e. SI (U) and SI~(D). Consequently both (U) electrons contribute to SI~(U) channel, whereas only one electron contributes to SI~(D) channel.

Now, our objective is to explore to what extent one may mimic the double ionization of a three-electron atom with a two-active electrons model. To that end, we consider the two scenarios introduced above. The presence of a third electron is taken into account by two factors: by the nuclear charge $Z=2$ and by considering spin degrees of freedom of the two-active electrons. The latter are deduced from the underlying ionization scenarios in the three-active electrons model. Namely, 
{in scenario 1, a  (U) electron is ionized first, followed by (U) or (D). Thus there are two possibilities, where either 
a (UU) or a (UD) pair is extracted in the double ionization event. In scenario 2, by contrast, the first electron is the
(D) electron, which leaves two equivalent U electrons for the second ionization step, so that the two-electron ionization always results in a (DU) pair.} Note, that (UD) and (DU) pairs are indistinguishable in a two-electron system. Each spin pair induces a wave function with a proper spatial symmetry --- (UU) indicates a problem with an antisymmetric wave function, whereas (UD) and (DU) refer to a spatially symmetric wave function. This is why the two different two-active electrons models, denoted by scenarios 3 and 4 in Fig.~\ref{fig1}, corresponding to (UU) and (UD) pairs are to be considered. We will refer to them as the antisymmetric and symmetric models, respectively.

For consistency, the DI potentials for the two electron models are imposed to be equal to the one from the three-electron model. The decision of keeping the DI potential values the same for both scenarios 3 and 4 is dictated by the fact that the DI potential has a strong effect on the ionization yield. At the same time ground state energies for symmetric and antisymmetric models are known to be quite different -- recall those of para- and ortho-helium \cite{Ruiz03,eckhardt2008}, for instance. The softening parameters for antisymmetric and symmetric models used to achieve our goal are $\epsilon_{antisym}=\sqrt{0.68}$ a.u. and  $\epsilon_{sym}=\sqrt{1.18}$ a.u. correspondingly.
The corresponding values of ionization potentials for both models are shown in the scenarios 3 and 4 in Fig. \ref{fig1}.
 
It is important to note that SI potentials are different for different scenarios. In the three-active electrons model SI potential is either 0.39 a.u. (scenario 1, when (U) electron is first ionized) and 1.21 a.u. (scenario 2, when (D) electron is first ionized). This should not be a surprise, as the remaining single ion has either two electrons with opposite spins or two electrons with the same spins. Thus corresponding wave functions are spatially symmetric or antisymmetric, and may be related to the ground and the exited state of the single ion, respectively. There is a good correspondence between three-active and two-active electrons models, with respect to SI potentials, when it is assumed that the first two escaping electrons both have spin up -- see scenarios 1 and 3 in Fig.~\ref{fig1}. In those two cases SI potentials are very close to each other (i.e. 0.39 a.u. and 0.35 a.u. in scenario 1 and 3, respectively). That correspondence suggests that the ionization occurs in the similar way. And this appears to be the case, as we will discuss in Sec.~\ref{sec:results_3e}. Note, however, that the antisymmetric model fits to one path realized in scenario 1, i.e. to (UU) only. The difficult part comes when one considers the first two escaping electrons having opposite spins. In such a case, two-electron model cannot distinguish between (UD) and (DU) channels, whereas three-active electrons model allows for the separation of them ((UD) is included in scenario 1, and (DU) -- in scenario 2). Consequently, the SI potential in scenario 4 is different from the SI potential in the first and second
scenario. Fortunately, its value lies approximately in the middle between those of scenario 1 and scenario 2, and is equal to 0.72 a.u. Let us note that scenarios 1 and 2 include quantum trajectories that end in the same state, namely, double ion with (U) electron. One could expect an interference between the (UD) and (DU) paths. Such an interference is an intrinsic element of the symmetric model. 

Any improvements of potential (\ref{ham2e}) by splitting, for example, the softening parameter into two, one corresponding to nuclear-electronic and the other to electronic-electronic interaction terms (like discussed in \cite{koval2007nonsequential}), for tuning SI potential values are not helpful. Such a manipulation cannot change symmetry properties of scenario 3 and scenario 4 that follow solely from their two-electron origin.

	\section{\label{sec:results_3e}Results and discussion}
	
\subsection{Preliminaries}	
In Fig.~\ref{fig2} and Fig.~\ref{fig3} we show the results of simulations obtained within the three-active and two-active electrons models, respectively -- single and double ionization yields are plotted as functions of the field amplitude. Note that respective ionization yields are split into various channels. To discern these channels we use a spatial criterion based on a division of space into regions corresponding to atom, single and double ions (and triple ions for the three-active electrons case), and calculating probability fluxes through boundaries between these regions~\cite{Thiede18,prauzner2008quantum,prauzner2007time}. 

The spatial criterion in our model allows one to distinguish between the two types of double ionization: the simultaneous, instantaneous escape of electrons and time delayed processes. The former is called a recollision-impact ionization (RII), known also as an electron-impact ionization (EII) or a recollision induced direct ionization (REDI)~\cite{weber2000correlated,weber00b,moshammer00}. The latter includes a variety of mechanisms such as the sequential double ionization (SDI) and recollision excitation with a subsequent ionization (RESI)~\cite{Shaaran11,Feuerstein01,deJesus04,Faria11}; another recently proposed time-delayed mechanism to be noted is ``slingshot nonsequantial double ionization'' \cite{Katsoulis18}. The channels distinguished by our space criteria are denoted RII and time delayed ionization (TDI), respectively. TDI in our model puts together SDI and RESI yields despite RESI being physically a non-sequential process. This is a drawback of our approach. As it is well known the importance of RESI manifests itself by the existence of the ``knee''. It is clearly visible in our TDI channels in Fig.~\ref{fig3}, while for a truly sequential channel such a ``knee'' should not appear.

For the purpose of expressing the electrons ionization order in time delayed processes we use notation (0-D-U) introduced previously in~\cite{Thiede18}: from neutral atom ``0'' the first (D) electron escapes, then does the second (U). For RII processes the notation without accounting for ionization order is used: (0-DU).

In the following we will compare results obtained with the use of three-active and two-active electrons models. Unfortunately, a direct comparison of ionization yields obtained within different models is not fruitful as models act in different restricted geometries. In such a case we may analyze instead  slopes, trends and overall shape of the curves rather than  the numerical values obtained. Whenever we compare different ionization channels within the same model we can additionally compare their magnitudes. We organize the section as follows: first we discuss the three-electron model results, then proceed with describing the two-electron model results comparing them with the three-electron model results.


	\subsection{Three-active electrons model results}

\begin{figure}
	\includegraphics[width=1.0\linewidth]{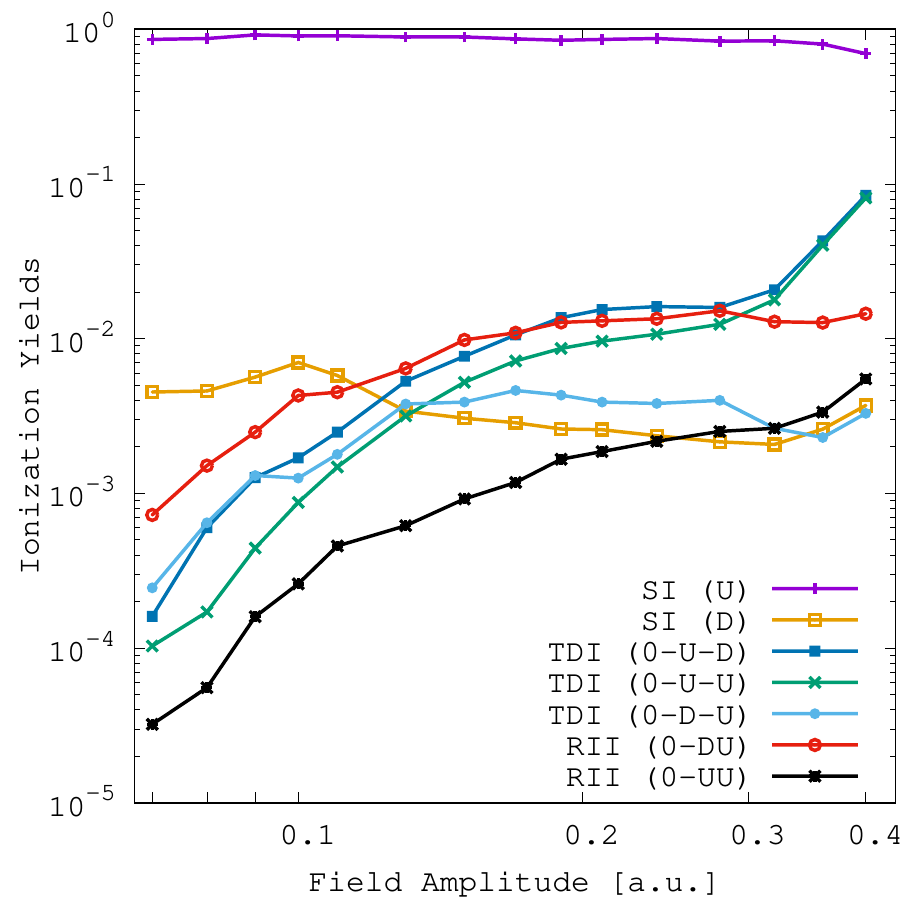}
	\caption{(Color online) Ionization yields as a function of electric field amplitude resolved for different ionization channels of the three-active electrons model (\ref{ham3e}). The channels are denoted as single ionization (SI), time delayed ionization (TDI) and recollision-impact ionization (RII) with the corresponding spin sequence of ionizing electrons placed in parentheses. 5-cycles long $\sin^2$-shaped pulse of frequency 0.06 a.u. (Eq.~(\ref{pulse})) has been used for simulations.}
	\label{fig2}
	\includegraphics[width=1.0\linewidth]{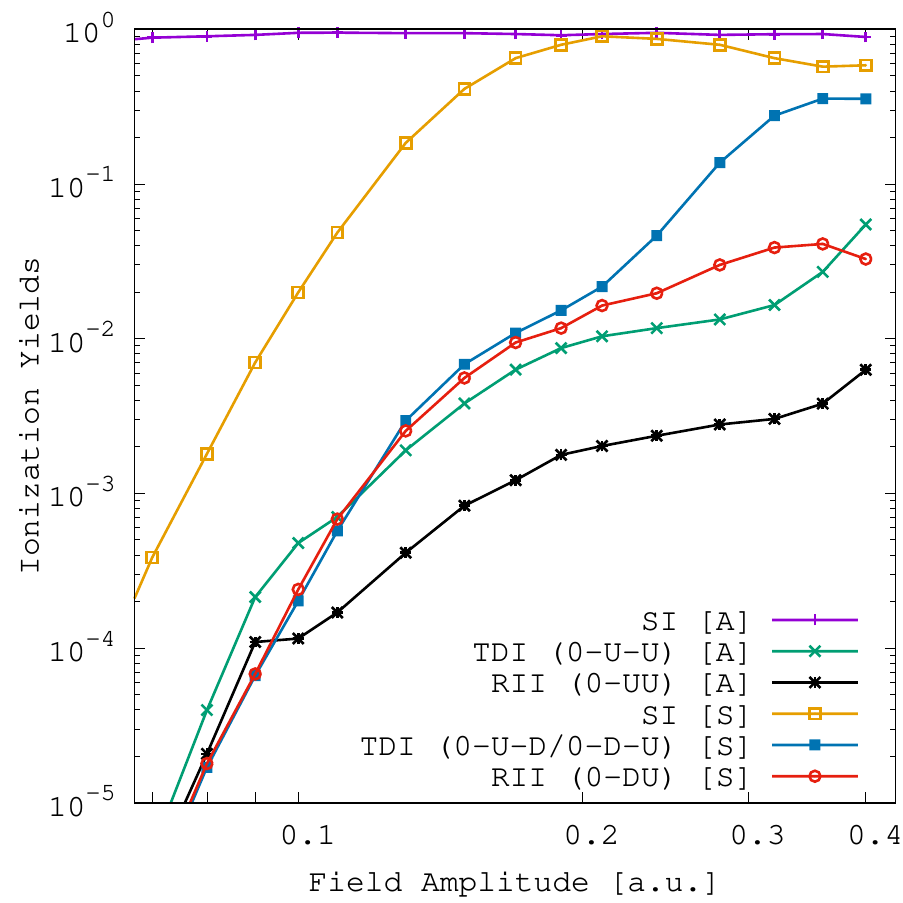}
	\caption{(Color online) Ionization yields vs electric field amplitude resolved for different ionization channels of two-electron atom (\ref{ham2e}) in both symmetric and antisymmetric configurations. Notation follows  Fig.~\ref{fig2}. Data corresponding to the symmetric and antisymmetric models are labeled [S] and [A], respectively.
	}
	\label{fig3}
\end{figure}

\paragraph{Single ionization.}
First we analyze single ionization yields. In the three-active electrons model it is clearly seen that the magnitude of the SI signal depends on the spin of ionized electron (compare SI (U) and SI (D) curves in Fig.~\ref{fig2}). The SI (U) signal evidently dominates over SI (D). While  two (U) electrons contribute to SI (U) yield (as compared to a single one for (D) yield) the origin of the observed large difference in yields can be traced to vastly different single ionization potentials in both cases. As seen from Fig.~\ref{fig1}, for the U electron $I_U=0.39$ a.u. while for the D electron $I_D=1.21$ a.u.  Using perturbative ideas one estimates that, for $\omega=0.06$ a.u., a U electron needs to absorb about 7 photons while the D electron needs as many as 21 photons! Another (less important) process leading to the small SI (D) signal is the depletion of singly charged ions due to a subsequent ionization from an antisymmetric state -- observe that after a maximum around $F_0=0.1$ a.u. the SI(D) signal remains well below $10^{-2}$ in Fig.~\ref{fig2}.

\paragraph{Double ionization -- TDI channel.}
An interesting observation within the discussed models is the fact that the three-active electrons model allows to distinguish between the process in which a (U) electron is ionized first followed by (D) electron from the process in which they ionize in a reverse order. The two processes are inherently indistinguishable in the two-active electrons model and are embedded in the symmetric model only. The remaining channel, i.e. (0-U-U), can be analyzed in both three-active and two-active electrons (antisymmetric) models.

As evident from Fig.~\ref{fig2}, ionization yields for TDI (0-U-D) and (0-D-U) show quite a different behavior. To some degree, this again reflects the different SI potentials for (U) and (D) electrons, while DI potentials are the same. For time delayed ionization one needs to consider both SDI and RESI. While for higher fields, above $F_0=0.3$ a.u., the SDI process starts to become significant, for lower amplitudes channels recollisional excitations dominate. The first step of RESI, single ionization, is more effective in the (0-U-D) channel as
 the TDI (0-U-D) signal clearly dominates over TDI (0-D-U) as the field amplitude increases -- see Fig.~\ref{fig2}. On the other hand, one can argue that more efficient ionization of the first (U) electron should lead to its more efficient rescattering and a higher excitation of the parent ion and thus should finally result in higher RESI when compared to the channel with the (D) electron being ionized first. However, this reasoning fails to describe trends observed for both signals in the low field amplitude region $F<0.1$ a.u. There, the (0-D-U) 
signal is comparable with (0-U-D). It has been suggested that, at low fields, RESI involves doubly excited states~\citep{Liao2017}. If that was the case, then both channels would become equivalent as the doubly excited states, due to their energy, should decay similarly in both channels. 
Thus our results, while not providing a proof of the claims in \citep{Liao2017}, at least are consistent with them.

Next, we compare TDI (0-U-D) and TDI (0-U-U) signals -- see Fig.~\ref{fig2}: both channels correspond to the same scenario 1 in Fig.~\ref{fig1}, but their yields are slightly different, especially for small field amplitudes. For SDI channel the order of electrons ionization should not matter at all; and indeed, for field amplitudes above $F\approx0.3$ a.u., when SDI plays important role, the two curves almost coincide. For less intensive fields recollisions have decisive impact on double ionization. The nearly parallel behavior of both signals in Fig.~\ref{fig2} suggests that the difference between ionization yields results from  field intensity-independent parameters, i. e. scattering cross sections of (D) and (U) electrons. Such a guess is reasonable since the spatial profiles of Ne$^{+}$ bound eigenstates depend on whether the ion contains same-spin or different spin electrons.  


Within the discussion above it is assumed that the recolliding electron is at the same time the electron that ionizes first. This assumption is not obvious, since the scenario when the striking electron can be captured by the ion while transferring energy for a release of a previously bound electron is possible. Still we suppose the latter option to be less probable because the recollision cross section is proportional to the ion's ionization cross section \cite{Landau3quantum} which drops with increasing of energy the bound electron gets during ionization. In the  extreme case when the rescattering electron has large energy and transfers most of it to the bound electron, the rescattering cross section is reduced to the classical Rutherford formula \cite{Landau3quantum} and thus is proportional to $1/E^2$, where $E$ is the energy gained by the bound electron. So, it is more likely for the rescattering electron to remain free rather than to be captured by the ion. The direct proof of the suggestion in our quantum simulation is not possible though. Our assumption may be partially validated by the classical trajectory method that allows one to track swapping of electrons during recollisions; Haan et al \cite{Haan07} estimated the number of double ionization events with electron swapping at $30\%$ within their classical model using similar field parameters (780 nm of wavelength, $F=0.1$ a.u. and 10 cycles of pulse duration that is twice longer than we have).

\paragraph{Double ionization -- RII channel.}
Let us now consider direct electron escapes. RII signal for  electrons escaping with opposite spins, i.e. the channel (0-DU), is always significantly larger than for electrons escaping with the same spin, i.e. the channel (0-UU) -- compare Fig.~\ref{fig2}. 
The ponderomotive energy needed for RII (UU) is $U_p=0.54$ a.u. that corresponds to $F=0.045$ a.u., while for RII (DU) $U_p=0.28$ a.u. with $F=0.03$ a.u.

Comparison of RII and TDI signals shows that the signal in the RII (0-DU) dominates other double ionization signals for amplitudes up to $F_0\approx 0.15$ a.u. Just below $F_0=0.2$ a.u. TDI (0-U-D) becomes comparable with RII (0-DU) and for amplitudes $F_ 0>0.3$ a.u. exceeds it, i.e. SDI becomes the dominant mechanism for double ionization. Also TDI (0-U-U) is greater than RII (0-DU) in that range of field amplitudes. On the contrary, the RII (0-UU) is negligible when compared with other signals -- this is again a manifestation of the suppression of a correlated escape due to the symmetry of the initial wave function. Only for large field amplitudes  does RII (0-UU) become comparable with TDI (0-D-U).

In general, the difference between RII and RESI channels is that double ionization in RII is localized in a small region near the origin of coordinates in our multielectron space, while RESI is not. The wave function for the antisymmetric configuration ($\psi(r_i,r_j)=-\psi(r_j,r_i)$) corresponding to (0-UU) channel has a nodal line along $r_i=r_j$ that is close to the direction of the correlated electronic escape during RII. This causes low probability of RII (0-UU) in comparison to that of  RESI and RII (0-DU). Another factor explaining high efficiency of (0-DU) channel is the unusually small value of 0.89 a.u. for the second electron ionization potential versus the single ionization potential $I=1.21$ a.u. in scenario 2: the rescattering electron has a high chance to directly ionize the parent ion rather than excite it. Such a property can be attributed to 3-electron configuration, because of spin degrees of freedom (see the discussion on ionization scenarios in Sec.~\ref{sec:pathways}), as the second electron ionization potential 
can never be smaller than the single ionization potential in any 2-electron model.

\paragraph{Double ionization -- final remarks.}
 Besides involving the dynamics of different spin-resolved channels, an overall enhancement of the double ionization signal for low field amplitudes in the three-active electrons model might be attributed to electron-electron Coulomb interactions that make rescattering cross section of the returning electron larger. For double ionization/excitation the third electron could serve like a catalyst in the small field regime. Such mechanism would be analogous to that responsible to a formation of ``knee'' as discussed by Szymanowski et al. \cite{Eberly00}. However, there is no easy way to support such a conclusion in a present study.

\begin{figure}
	\includegraphics[width=1.0\linewidth]{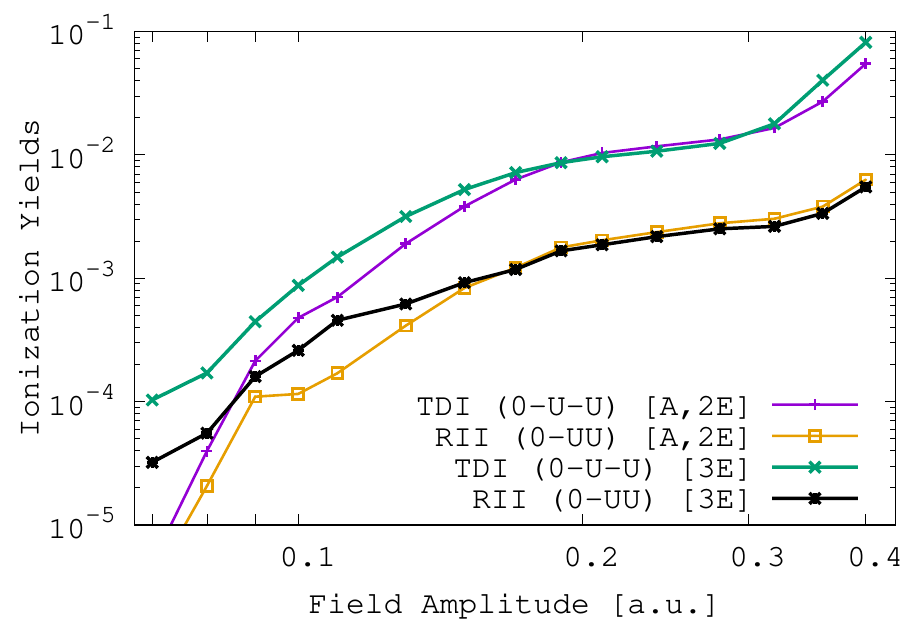}
	\caption{(Color online) Ionization yields vs electric field amplitude: selected curves from Figs. \ref{fig2} and \ref{fig3}.}
	\label{fig6}
\end{figure}

	\subsection{Comparison with two-active electrons model results}
	
\begin{figure}
	\includegraphics[width=1.0\linewidth]{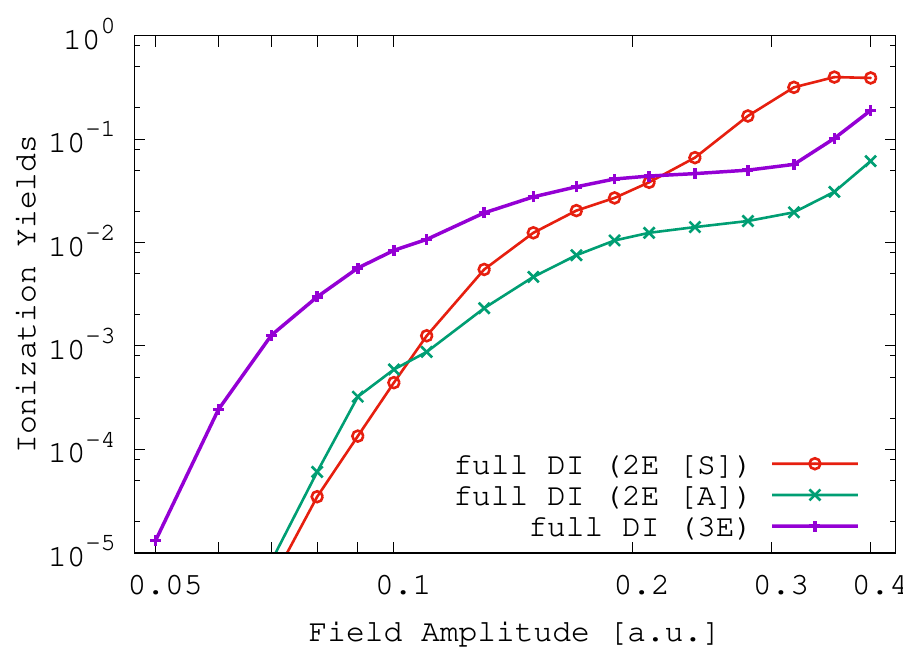}
	\caption{(Color online) Ionization yields vs electric field amplitude: selected curves from Figs. \ref{fig2} and \ref{fig3}. The ``full DI'' curves were obtained by summing all possible double ionization channels.}
	\label{fig4}
\end{figure}

Comparison of curves corresponding to the same ionization channels in three-electron model (Fig.~\ref{fig2}) and two-electron models (Fig.~\ref{fig3}) implies that the symmetric model data differ in shape considerably from their three-electron counterparts. On the contrary, the antisymmetric model data, essentially, follow their analogues (see Fig. \ref{fig4}).  Nevertheless, we briefly review data obtained with both models contrasting them to proper channels in the three-electron model. 

\paragraph{Antisymmetric model discussion.}
 In the antisymmetric two-active electrons model (see Fig.~\ref{fig3}) there are only (U) electrons, therefore its SI signal corresponds solely to the SI(U) channel in the three-active electrons model. So it is of no wonder that both the SI(U) signal of the three-active electrons model and the SI signal of the antisymmetric model exhibit an almost complete saturation of the yield in the range of field amplitudes considered. 
 
TDI signal obtained within antisymmetric model may be compared to TDI (0-U-U) signal of the three-electron model only -see Fig.~\ref{fig6}.
Generally, 
we conclude that both signals have similar shapes. 
For lower field amplitudes, however, the TDI curve in the antisymmetric two-electron model is slightly steeper than its counterpart in the other model.
 

The results of simulations for RII channel (0-UU) in three- and two- active electron models, 
are also compared in Fig.~\ref{fig6}. A very good agreement between the models is found for large field amplitudes $F_0>0.15$. For lower amplitudes, while both curves are still parallel to each other, the RII from the three-electron case starts to grow earlier. Such a feature might suggest the importance of the three-electron correlations.


\paragraph{Symmetric model discussion.}
The SI channel in the symmetric model corresponds partially to the SI(U) and partially to the SI(D) channels of the three-electron model, but the SI[S] signal does not match neither SI(U) nor SI(D). Rather, it shows the shape known from earlier studies of two-electron models (see for example~\citep{prauzner2007time,Lappas_1998,PhysRevLett.83.520}).
Clearly even a single electron physics of the three-electron atom cannot be simulated within the restricted two-electron symmetric model.


In general, double ionization signals for the symmetric model grow more rapidly with a field amplitude than the corresponding signals for the three-electron model, regardless the channel. Let us compare TDI (0-U-D) and (0-D-U) from Fig.~\ref{fig2} with TDI[S] from Fig.~\ref{fig3}. For field amplitudes larger than $F_0=0.2$ a.u. TDI[S] curve has larger slope than either of its counterparts. In the symmetric model, in that range of amplitudes SDI becomes an important ingredient in the TDI signal because less energy is required to surmount the second ionization potential. In the three-active electrons model such an increase is {not observed until the field amplitudes exceeds  $F_0=0.3$ a.u., and then for the TDI (0-U-D) channel only.}

	\section{\label{sec:conclusion}Conclusions}

We have studied double ionization of three-electron atom within a simplified three-active electrons model. We have compared single and double ionization yields 
with those 
obtained with the use of judiciously chosen two-active electrons models. The two-active electrons models have been designed to include spin degrees of freedom by a proper choice of symmetry of initial wave functions. In each of the analyzed models we investigated signals of different spin ionization channels. 
{It is currently not possible to experimentally resolve these channels, but as our results show
they have effects on }
(i) absolute values of double ionization (normalized to single ionization yield, for example) and (ii) of knee slope position.

Electronic correlations can affect double ionization because of Coulomb interactions and spin configurations. The former might be partially responsible for the shift of the double ionization knee slope to lower fields. The latter defines peculiarities of each ionization channel yields dependencies on field amplitude and, thus, 
total double ionization yield. The influence of Coulomb electron-electron interactions on the double ionization while being intriguing appeared to elude the research method used. 

Comparison of the different cases shows that double ionization in the case of a correlated three-electron system can not, in principle, be properly represented by a set of two-electron subsystems. This results in a considerable limitation of precision and applicability of any two-electron model used for simulating three- and presumably other multi-electron atoms double ionization. 
Taking into account wavefunction symmetry properties resulting from spin considerations allows to define different ionization channels. The differences in  shapes of ionization curves for  those channels can be used to differentiate between them. This is in contrast to the situation in the two-electron atom case, 
where such differences can not be resolved.

From our analysis it follows that two-electron antisymmetric model is in better qualitative agreement with double ionization of correlated three-electron atoms  than the symmetric one. Curiously, antisymmetric model shares a property with the celebrated Rochester model \cite{haan2002}: both prevent simultaneous escape of electrons, however, due to 
different reasons. In the first case, the simultaneous double ionization suppression is caused by the wave function symmetry (spin) depleting the area around $r_1=r_2$, 
while in the second case it is the overestimated Coulomb repulsion between electrons that restricts electrons from approaching each other when they are away from nucleus.
 
Despite application of restricted dimensionality models in this work, the results obtained can be generalized to real three-dimensional atoms. The spin structure remains the same, so that the qualitative effects 
discussed here should be independent of space dimensions.
Our observations extend previous studies of triple ionization \cite{Ruiz05,Thiede18} and show that spin electron correlations can have significant effects on double ionization as well.

	\section{Acknowledgements}
A support by  PL-Grid Infrastructure is acknowledged.
This work was realized under   National Science Centre (Poland) project Symfonia  No. 2016/20/W/ST4/00314.

\appendix

\section{\label{app:wavefunction}Three-electron ground state wave function.} 
In a model atom with 3 active electrons one has to take care about spin degrees of freedom~\cite{Ruiz05}. For two-active electrons models it is typically assumed that electrons have opposite spins, and the spatial part of the ground state wave function is symmetric with respect to an exchange of particles. Now, when the third electron comes into play it is impossible to construct a wave function in a way that leaves its spatial part symmetric with respect to exchange of each pair of particles. For the sake of satisfying Pauli's exclusion principle the general form of electronic wave function can be written as \cite{Ruiz05}:
\begin{multline}
	\Psi_{\alpha \alpha \beta} (r_1,r_2,r_3,t) \propto \alpha(1)\alpha(2)\beta(3)\psi_{12} (r_1,r_2,r_3,t) \\
	+ \beta(1)\alpha(2)\alpha(3)\psi_{23} (r_1,r_2,r_3,t) \\
	+ \alpha(1)\beta(2)\alpha(3)\psi_{13} (r_1,r_2,r_3,t),
\end{multline}
where $\alpha(i)$ and $\beta(i)$ denote spin functions corresponding to spin-up and spin-down states, respectively. The spatial functions $\psi_{ij} (r_1,r_2,r_3,t)$ are antisymmetric in 
{the ($r_i,r_j$) plane and symmetric with respect to the third electron.}

The Hamiltonian (\ref{ham3e}) commutes with the electron exchange operators and, thus, the symmetry of the wave function is preserved during the time evolution. As a consequence, one can treat the evolution of each 
wave function $\psi_{ij}$ independently from the other terms. Importantly, it is sufficient to evolve only one of the $\psi_{ij}$'s --- results for the remaining two are found by simple permutations of the coordinates~\cite{Ruiz05}. 

\bibliography{reference}

\end{document}